\documentclass{iaus}
\usepackage{graphicx}

\title[Bar Lengths, Colours and Ages] 
{On the Lengths, Colours and Ages of Bars}

\author[Gadotti \& de Souza]   
{Dimitri A. Gadotti$^{1}$ \and Ronaldo E. de Souza$^2$}

\affiliation{$^1$Max-Planck-Institut f\"ur Astrophysik, Karl-Schwarzschild-Str. 1,
D-85741, Garching bei M\"unchen, Germany
\break email: dimitri@mpa-garching.mpg.de\\[\affilskip]
$^2$Departamento de Astronomia, Universidade de S\~ao Paulo, Rua do Mat\~ao, 1226, 05508-090,
S\~ao Paulo-SP, Brazil}

\pubyear{2006}
\volume{235}  
\pagerange{???--???}
\date{?? and in revised form ??}
\jname{Galaxy Evolution across the Hubble Time}
\editors{F. Combes \& J. Palous, eds.}
\begin{document}

\maketitle
\begin{abstract}
In an effort to obtain further observational evidences for secular evolution processes in
galaxies, as well as observational constraints to current theoretical models of secular
evolution, we have used BVRI and Ks images of a sample of 18 barred galaxies to measure
the lengths and colours of bars, create colour maps and estimate global colour gradients.
In addition, applying a method we developed in a previous article, we could distinguish
for 7 galaxies in our sample those whose bars have been recently formed from the ones with
already evolved bars. We estimated an average difference in the optical colours between
young and evolved bars that may be translated to an age difference of the order of 10 Gyr,
meaning that bars may be long standing structures. Moreover, our
results show that, on average, evolved bars are longer than young bars.
This seems to indicate that, during its evolution, a bar grows longer by capturing stars
from the disk, in agreement with recent numerical and analytical results.
\end{abstract}
\firstsection 
\section{Results and Implications}

The B-V colour difference between young and evolved bars is 0.4 mag, which can be translated
to an age difference of 10 Gyr. This means that bars can be robust structures,
in agreement with recent $n$-body simulations and observations of barred galaxies at
higher redshifts. The young bars in our sample have an average length of 5.4$\pm$1.6 Kpc,
while the evolved bars have an average length of 7.5$\pm$1.2 Kpc, consistent with
recent theoretical expectations that bars grow longer while aging.
Young bars are preferentially found in late-type spirals, indicating that
bar recurrence may be more frequent in gas-rich, disk-dominated galaxies. We also found that
AGN are preferentially hosted by galaxies with young bars, suggesting that the fueling of AGN
by bars happens in short timescales and that a clearer bar-AGN connection
would be found in a sample of galaxies with young bars. We have also found that bar colours might
be used as a proxy for bar ages. Enlarging the sample of bars with measured ages is paramount
to calibrate this relation, confirm these results, compare in more detail observations
and models, and better understand secular evolution. See \cite[Gadotti \& de Souza (2005,
2006)]{PaperI,PaperII} for further details.

\begin{acknowledgments}
DAG would like to thank the Deutsche Forschungsgemeinschaft and the Max-Planck-Gesellschaft
for financial support.
\end{acknowledgments}

\end{document}